\documentclass[showpacs,preprintnumbers,%
amsmath,amssymb,%
twocolumn
]{revtex4}
\usepackage[dvips]{graphicx}
\newcommand{\up}{\uparrow}
\newcommand{\dn}{\downarrow}
\newcommand{\sgn}{{\mathbf{sgn}}}
\newcommand{\La}{\Lambda}
\newcommand{\la}{\lambda}
\newcommand{\Ne}{{N_\mathrm{e}}}
\newcommand{\Neup}{{N_{\mathrm{e},\uparrow}}}
\newcommand{\Nedn}{{N_{\mathrm{e},\downarrow}}}
\newcommand{\PhiG}{\Phi_{\theta,N_\mathrm{e}}}

\newcommand{\qed}{\rule{6pt}{6pt}}
\newcommand{\calK}{\mathcal{K}}
\newcommand{\rmi}{\mathrm{i}}
\newcommand{\rme}{\mathrm{e}}
\newcommand{\sumsigma}{\sum_{\sigma=\uparrow,\downarrow}}
\newcommand{\cs}[1]{c_{#1,\sigma}}
\newcommand{\csd}[1]{c_{#1,\sigma}^\dagger}

\newcommand{\as}[1]{a_{#1,\sigma}}
\newcommand{\asd}[1]{a_{#1,\sigma}^\dagger}
\newcommand{\aup}[1]{a_{#1,\uparrow}}
\newcommand{\adn}[1]{a_{#1,\downarrow}}
\newcommand{\aupd}[1]{a_{#1,\uparrow}^\dagger}
\newcommand{\adnd}[1]{a_{#1,\downarrow}^\dagger}
\newcommand{\tas}[1]{\tilde{a}_{#1,\sigma}}
\newcommand{\tasd}[1]{\tilde{a}_{#1,\sigma}^\dagger}

\newcommand{\taupd}[1]{\tilde{a}_{#1,\up}^\dagger}
\newcommand{\tadnd}[1]{\tilde{a}_{#1,\dn}^\dagger}
\newcommand{\tasp}[1]{\tilde{a}_{#1,\sigma}^{\prime}}
\newcommand{\taspd}[1]{\tilde{a}_{#1,\sigma}^{\prime\dagger}}

\newcommand{\bs}[1]{b_{\theta,#1,\sigma}}
\newcommand{\bsd}[1]{b_{\theta,#1,\sigma}^\dagger}
\newcommand{\bupd}[1]{b_{\theta,#1,\uparrow}^\dagger}
\newcommand{\bdnd}[1]{b_{\theta,#1,\downarrow}^\dagger}
\newcommand{\bup}[1]{b_{\theta,#1,\uparrow}}
\newcommand{\bdn}[1]{b_{\theta,#1,\downarrow}}
\newcommand{\bsp}[1]{b_{\theta,#1,\sigma}^{\prime}}
\newcommand{\bspd}[1]{b_{\theta,#1,\sigma}^{\prime\dagger}}

\newcommand{\bms}[1]{b_{\theta,#1,-\sigma}}
\newcommand{\bmsd}[1]{b_{\theta,#1,-\sigma}^\dagger}
\newcommand{\bcs}[1]{\bar{c}_{#1,\sigma}}
\newcommand{\bcsd}[1]{\bar{c}_{#1,\sigma}^\dagger}
\newcommand{\bcup}[1]{\bar{c}_{#1,\uparrow}}
\newcommand{\bcdn}[1]{\bar{c}_{#1,\downarrow}}
\newcommand{\bcupd}[1]{\bar{c}_{#1,\uparrow}^\dagger}
\newcommand{\bcdnd}[1]{\bar{c}_{#1,\downarrow}^\dagger}
\newcommand{\Htheta}{H_\theta}
\newcommand{\Hhop}{H_{\mathrm{hop}}}
\newcommand{\Hint}{H_{\mathrm{int},\theta}}
\newcommand{\Gram}[1]{(G)_{#1}}
\newcommand{\rGram}[1]{(G^{-1})_{#1}}

\newcommand{\Uth}[1]{(U_\theta)_{#1}}

\newcommand{\zthd}{\zeta_{\theta}^\dagger}
\newcommand{\zthdn}{\left(\zeta_{\theta}^\dagger\right)^\frac{N_\mathrm{e}}{2}}

\begin{document}

\title{ 
 Extended Hubbard Model with Unconventional Pairing in Two Dimensions  
}
\author{Akinori Tanaka$^1$ and Masanori Yamanaka$^2$}
 \affiliation{%
 $^1$Department of Applied Quantum Physics, Kyushu University,
 Fukuoka 812-8581, Japan\\
 $^2$Department of Physics, College of Science and Technology, 
 Nihon University,
 Tokyo 101-8308, Japan}

\date{March 31, 2005}
\begin{abstract}
We rigorously prove that an extended Hubbard model with attraction
in two dimensions has an unconventional pairing
ground state for any electron filling.
The anisotropic spin-0 or anisotropic spin-1 
pairing symmetry is realized,
depending on a phase parameter characterizing
the type of local attractive interactions.
In both cases the ground state is unique.
It is also shown that in a special case,
where there are no electron hopping terms,
the ground state has Ising-type N\'eel order
at half-filling, when on-site repulsion is furthermore added.
Physical applications are mentioned.
\end{abstract}
\pacs{71.10.Fd, 74.20.-z}
\maketitle
Unconventional superconductivity with gap symmetries
other than the conventional $s$-wave has been found ubiquitously
in correlated electron systems.
Examples include heavy fermions~\cite{SU91}, 
high $T_\mathrm{c}$ cuprates~\cite{TK00},
ruthenate~\cite{MRS01}, organic conductors~\cite{Kanoda97}, etc. 
The common feature of those compounds 
is proximity of antiferromagnetic or ferromagnetic order.
A vast number of 
studies have been extensively devoted 
to revealing nature of these phenomena.
So far, within the mean-field approach,
it is recognized that 
effective electron-pair attraction
depending on electron momentum
can cause unconventional superconductivity, but, 
there still is no convincing evidence 
which model captures the mechanism truly. 
As a many-body problem, it is an extremely hard task
to make a definite criterion to distinguish 
the validity of the models
beyond the mean-field level.
Indeed, electron systems exhibit various physical phenomena, 
relying on a subtle interplay between kinetic and interaction energies. 
It is thus desirable to rigorously establish the occurrence of
unconventional pairing in concrete models of correlated electrons. 
The attempts in this direction
will shed light on the mechanism for unconventional superconductivity
and, in turn, give us useful information 
about possible sources of effective pair attraction 
in real materials.  

In this paper, we rigorously construct a series of the electronic models
having ground states with unconventional pairing symmetries. 
We consider a two dimensional tight-binding model 
with attractive interactions which act on electrons 
occupying certain localized single-electron states
(corresponding to \eqref{eq:a-operator} 
and \eqref{eq:b-operator} below).
For each even number of electrons,
the model is proved to have the unique ground state~\cite{comment4} 
in which all electrons form 
anisotropic pairs with spin 0 
or spin 1,
depending on a phase parameter ($\theta$ in \eqref{eq:b-operator})
of the localized states.
Here, we treat the model in two dimensions, 
since the case is most relevant to the experiments
mentioned above.
Extensions of the present method and idea to 
higher dimensional systems or other lattice structures
are straightforward.

Remarkably, unlike usual mean-field Hamiltonians,
our Hamiltonian conserves the electron number.
The occurrence of the electron-pair condensation 
is thus non-trivial in the present model.  
To our best knowledge, this is the first time
that a model Hamiltonian is proved to exhibit condensation
of unconventional electron pairs 
including spin-1 pairs~\cite{Tanaka03}.
Further advantage of our model is a relation to proximity 
of magnetic orders.
In a case, 
where there are no electron hopping terms,
the model exhibits antiferromagnetism at half-filling,
when on-site repulsion terms of Hubbard-type 
are furthermore introduced.
The model 
is expected to exhibit a quantum phase transition
between the superconducting and the antiferromagnetic states,
which is an essential feature of the cuprate superconductors,
when parameters are varied 
away {from} an exactly solvable point in a parameter space.

Let us define the model.
Let $\La$ be 
a rectangular lattice of the form $\La={[1,L_1]\times[1,L_2]\cap \Bbb{Z}^2}$
with periodic boundary conditions.  
It is assumed that $L_1$ is an odd positive integer and $L_2=L_1+2$.
(We need these conditions to prove the uniqueness of the ground state 
as we will show later.)
We denote by $\cs{x}$ ($\csd{x}$) the annihilation (creation) 
operator for an electron with spin $\sigma=\up,\dn$
at site $x$.
They satisfy the anticommutation relations 
$\{\csd{x}, c_{y,\tau}^\dagger\} = 
 \{\cs{x}, c_{y,\tau}\} = 0$
and 
$\{\csd{x},c_{y,\tau}\}=\delta_{x,y}\delta_{\sigma,\tau}$. 
We denote by $\Phi_0$ a state without electrons and by $\Ne$
the electron number. 
\begin{figure}[b]
\begin{center}
 \begin{tabular}{lcl}
  (a) &\hspace*{1cm} & (b)
 \\
 \raisebox{.2cm}{\includegraphics[width=.1\textwidth]{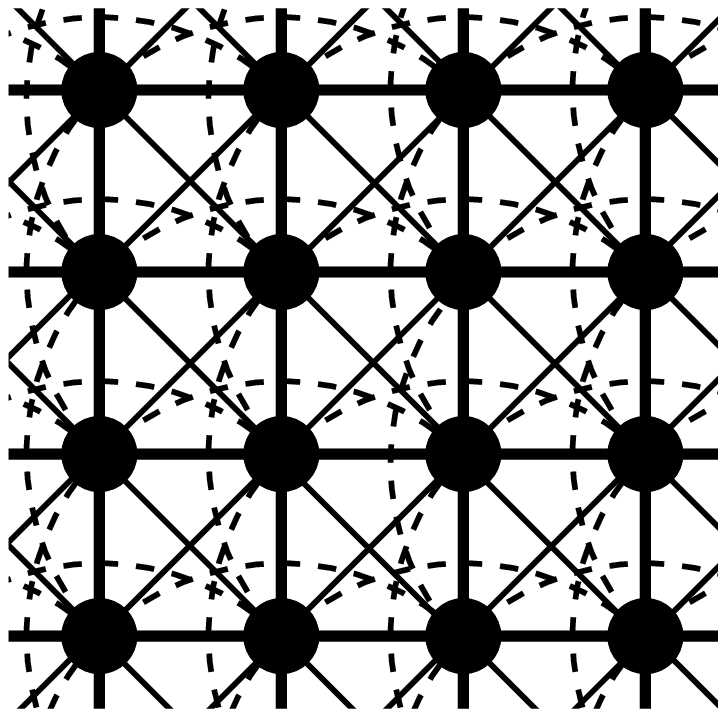}}
  &&
 \includegraphics[width=.2\textwidth]{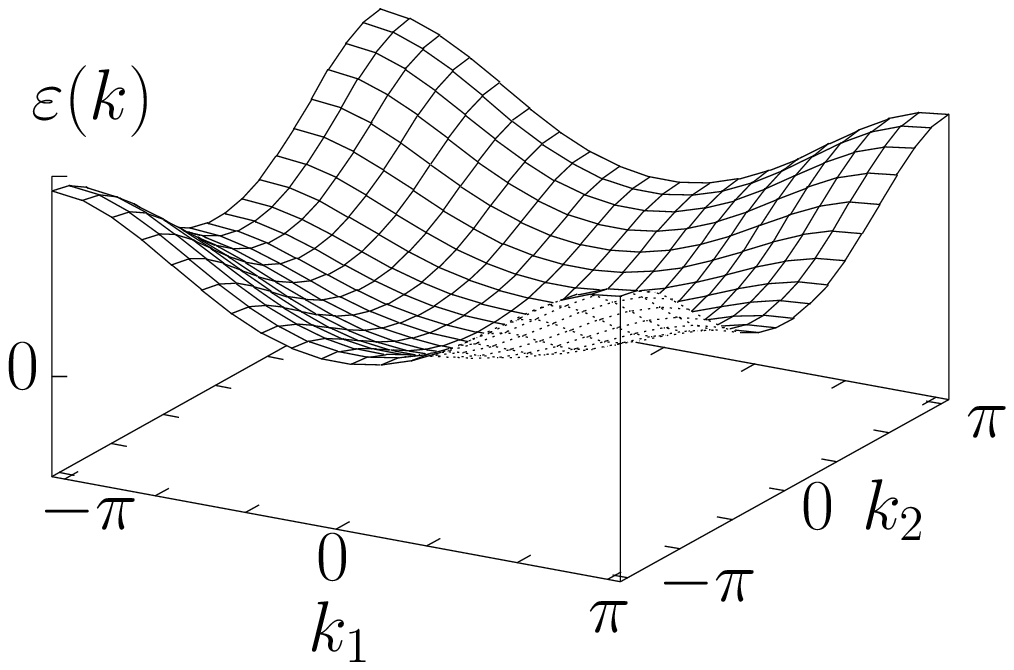}
 \end{tabular}
\end{center}
\caption{(a) Lattice structure. (b) Dispersion relation.}
\label{fig:lattice}
\end{figure}

The hopping part of our Hamiltonian is given by 
$
 \Hhop = \sum_{x,y\in\La}\sumsigma t_{x,y}\csd{x}\cs{y}
$
where $t_{x,y}=(1+4\la^2)t$ if $x=y$, $t_{x,y}=-2\la t$ if $|x-y|=1$,
$t_{x,y}=2\la^2t$, if $|x-y|=\sqrt{2}$, $t_{x,y}=\la^2t$ if $|x-y|=2$,
and zero otherwise~\cite{comment1}.
Here, it is assumed that $t>0$ and $-1/4<\la<1/4$. 
In the wave space, it is represented as
$
 \Hhop = \sum_{k\in \calK}\sumsigma\varepsilon(k)\bcsd{k}\bcs{k}
$
where 
$\varepsilon(k)=tg^2(k)$ with $g(k)=1-2\la\cos{k_1} - 2\la\cos{k_2}$ for 
$k=(k_1,k_2)$,
$\bcs{k}=1/\sqrt{|\La|}\sum_{x\in\La} \rme^{\rmi k\cdot x}\cs{x}$,
and 
\begin{equation}
 \calK=
  \left\{
   (\frac{2{\pi}n_1}{L_1},\frac{2{\pi}n_2}{L_2})~|~
   n_l=0,\pm1,\dots,\pm\frac{L_l-1}{2}
  \right\} .
\end{equation}
The lattice structure and the single-electron dispersion
relation are shown in Figs.~\ref{fig:lattice}~(a) and (b).

Let us introduce new fermion operators
corresponding to the single-electron states localized 
in the vicinity of site $x\in\La$ as follows:
\begin{eqnarray}
\label{eq:a-operator}
 \as{x}  =  \cs{x} - \la\sum_{y\in \La;|y-x|=1} \cs{y},\\ 
\label{eq:b-operator}
 \bs{x}  =  \sum_{y\in \La;|y-x|=1} \rme^{-\rmi\theta\cdot(y-x)}\cs{y}
\end{eqnarray}
with $\theta\in\{\alpha,\beta,\gamma\}$ where 
$\alpha=(0,0), \beta=(0,\pi)$ and $\gamma=(\pi/2,\pi/2)$.  
The interaction discussed in this paper
is attraction between electrons 
in these localized states. 
The interaction part of our Hamiltonian is given by
\begin{equation}
\label{eq:interaction}
 \Hint = -W \sum_{x\in\La}\sumsigma 
        \bmsd{x}\bms{x}\asd{x}\as{x} 
\end{equation}
with $W>0$.
One easily finds that 
the summand in \eqref{eq:interaction}
is bounded below by $-4(1+4\la^2)W$,
which is attained by the states of the form $\asd{x}\bmsd{x}\cdots\Phi_0$.  
This indicates that
$\Hint$ describes
attraction between two electrons with opposite spins.

The whole Hamiltonian of our model is given by
\begin{equation}
\label{eq:Hamiltonian-1}
\Htheta=\Hhop+\Hint+
 v_\theta\sumsigma\bcsd{0}\bcs{0}
\end{equation}
where $v_\theta=0$ if $\theta=\alpha$ and $v_\theta>0$ otherwise.
The last term is added for a technical reason
to show the uniqueness.

To state our main result, we need to introduce further notation.
Let $G$ be the Gram matrix for the $a$-operator~\eqref{eq:a-operator} 
whose matrix elements
are given by $\Gram{x,y}=\{\asd{x},\as{y}\}$. 
By a straightforward calculation, one finds that $G$ 
is regular and that its inverse matrix is given by 
$\rGram{x,y}=1/|\La|\sum_{k\in\calK}g^{-2}(k)\rme^{\rmi k\cdot(x-y)}$.  
Thus, it is possible to define dual operators of the $a$-operator as
$\tas{x}=\sum_{y\in\La}\rGram{y,x}\as{y}$, which satisfy
\begin{equation}
\label{eq:anticommutation-a} 
\{\asd{x},\tilde{a}_{y,\tau}\}=
\{\tasd{x}, a_{y,\tau}\}
=\delta_{x,y}\delta_{\sigma,\tau}.
\end{equation}
Since $\{\tasd{x}\Phi_0\}_{x\in\La}$ spans the single-electron Hilbert
space, the $b_\theta$-operators~\eqref{eq:b-operator} are expanded as
\begin{equation}
\label{eq:expansion-b}
 \bs{x} = \sum_{y\in\La} \Uth{y,x}\tas{y}.
\end{equation}
Here, the expansion coefficients $\Uth{y,x}$ are
given by 
$\Uth{y,x}=\{\asd{y},\bs{x}\}$.
One finds that
$\Uth{x,y}=\Uth{y,x}=(U_\theta)_{y,x}^\ast$ for $\theta=\alpha,\beta$
while $\Uth{x,y}=-\Uth{y,x}=(U_{\theta})_{y,x}^\ast$ for $\theta=\gamma$.

Using $\Uth{x,y}$, let us define 
\begin{equation}
\label{eq:zthd}
 \zthd = \sum_{x,y\in\La}\Uth{x,y}\taupd{x}\tadnd{y},
\end{equation}
which are the creation operators for electron-pairing states.
The main result in this paper is as follows:

\textit{Theorem. Suppose $\lambda\ne0$ and 
consider $\Htheta$ with $W=t/4$ and fixed $\Ne$ less
than $2|\La|$.
When $\Ne$ is even, the ground state is unique, has zero energy,
and is given by
\begin{equation}
\label{eq:ground-state}
 \PhiG=\zthdn\Phi_0.
\end{equation}
For odd $\Ne$ the ground state has positive energy.}

Before proceeding to the proof, 
we discuss the properties of $\Hint$ and pairing states. 
\begin{figure}
\begin{tabular}{lll}
(a) & (b) & (c)
\\
 \includegraphics[width=.15\textwidth]{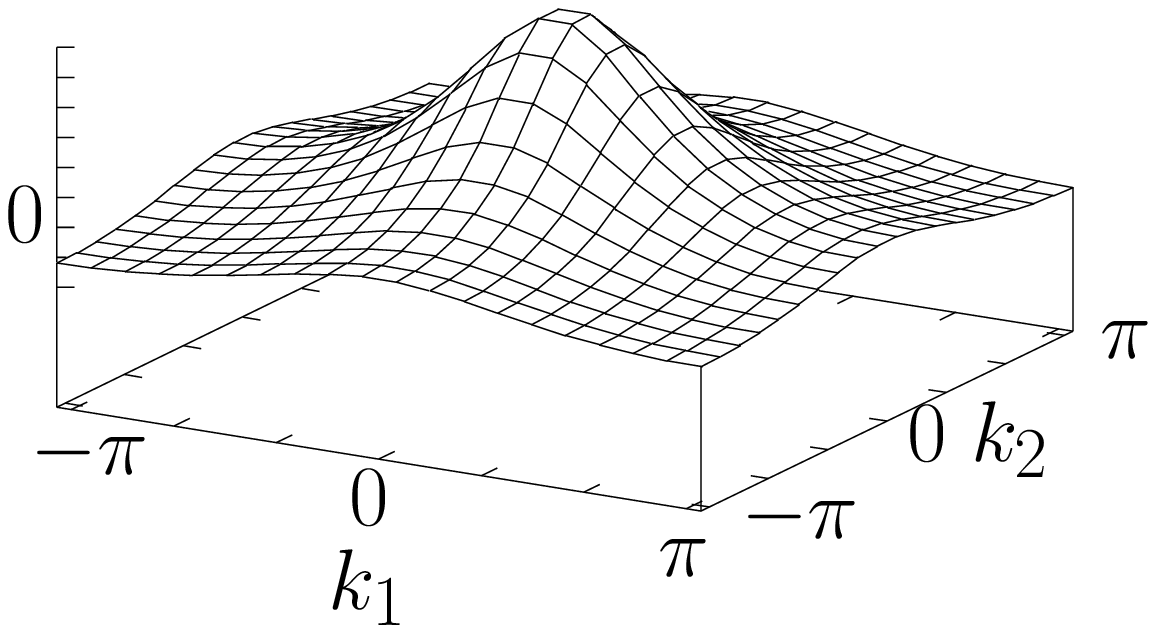}&
 \includegraphics[width=.15\textwidth]{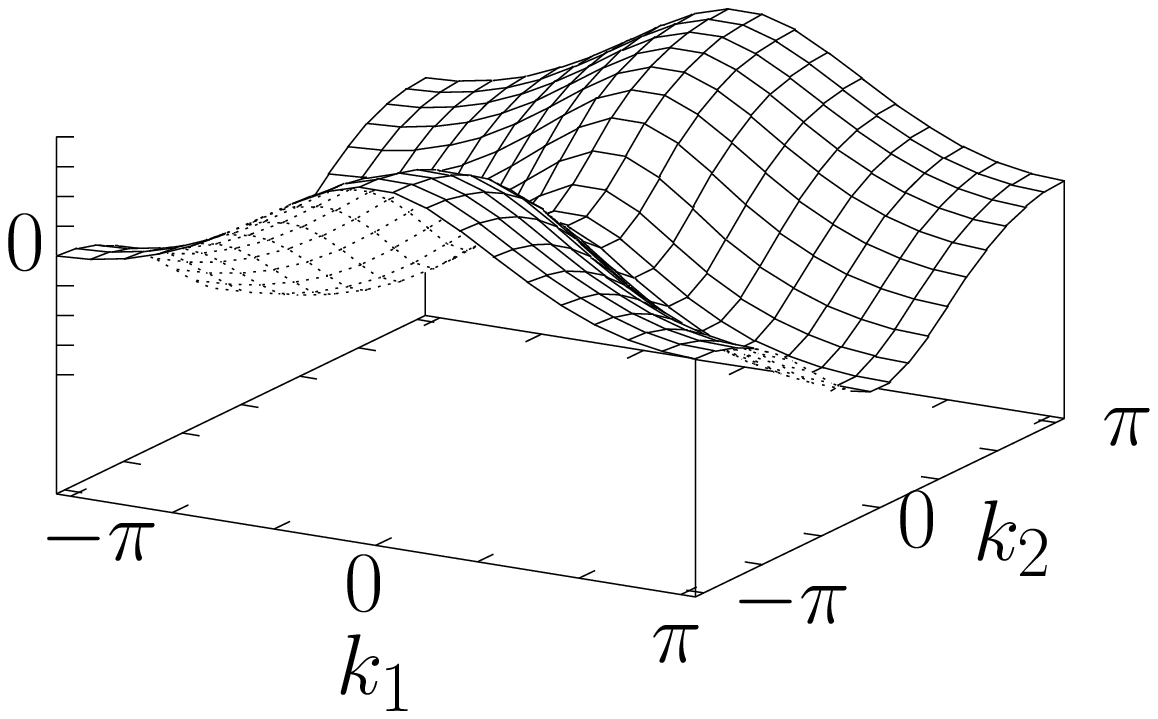}&
 \includegraphics[width=.15\textwidth]{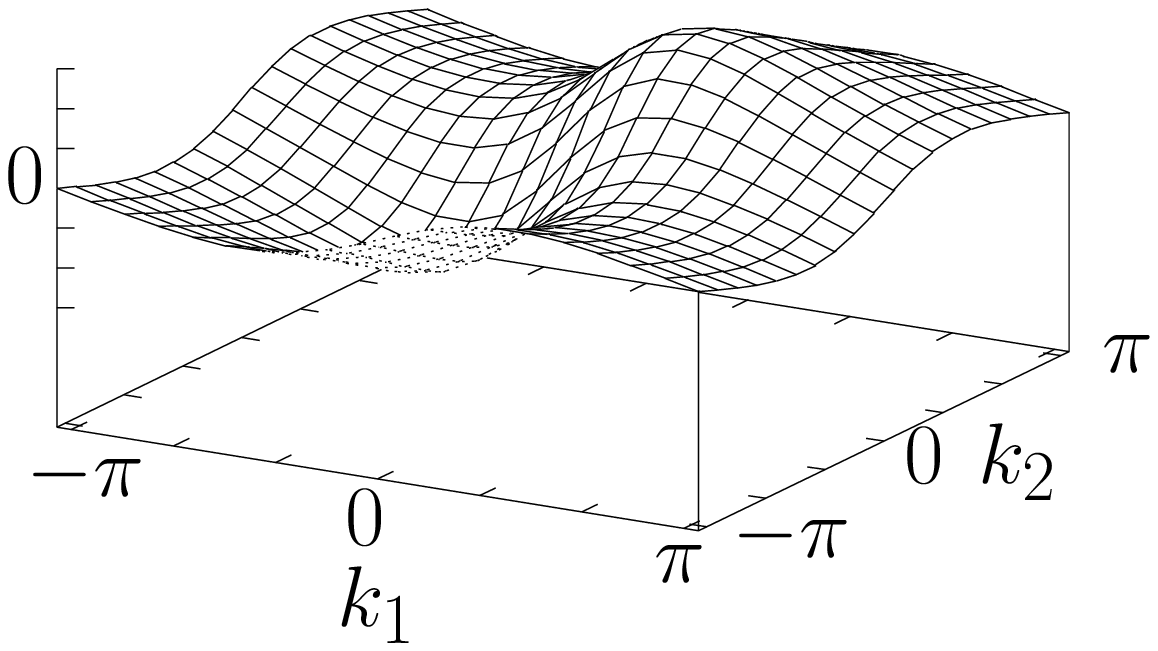}
\end{tabular}
\caption{Wave-vector dependence of $g_\theta(-k)/g(k)$. 
         (a) $\theta=\alpha=(0,0)$. (b) $\theta=\beta=(0,\pi)$. 
          (c) $\theta=\gamma=(\pi/2,\pi/2)$.}  
\label{fig:g-theta}
\end{figure}

By using the Fourier transforms of the $c$-operator, 
the fermion operators $\asd{x}$, $\bsd{x}$ are expanded as
\begin{eqnarray}
\label{eq:ax}
 \asd{x} = \frac{1}{\sqrt{|\La|}}\sum_{k\in\calK}
           g(k)\rme^{\rmi k\cdot x}\bcsd{k},\\
\label{eq:bx}
 \bsd{x} = \frac{1}{\sqrt{|\La|}}\sum_{k\in\calK}
           g_\theta(k)\rme^{\rmi k\cdot x}\bcsd{k},
\end{eqnarray}
where 
$g_\theta(k)=2(\cos(k_1+\theta_2)+\cos(k_2+\theta_2))$
for $k=(k_1,k_2)$ and $\theta=(\theta_1,\theta_2)$.
One also finds {from} \eqref{eq:ax} that 
$\tasd{x}=1/\sqrt{|\La|}\sum_{k\in\calK}
           g^{-1}(k)\rme^{\rmi k\cdot x}\bcsd{k}$.
Substitution of this and \eqref{eq:bx} into 
$\zthd=\sum_{x\in\La}\taupd{x}\bdnd{x}$, 
which follows {from} \eqref{eq:expansion-b} and \eqref{eq:zthd},
yields
\begin{equation}
 \zthd=\sum_{k\in\calK}\frac{g_\theta(-k)}{g(k)}\bcupd{k}\bcdnd{-k}.
\end{equation}
The precise expressions of $g_\theta(k)$ are given by
$g_{\alpha}(k)=2(\cos{k_1}+\cos{k_2})$,
$g_{\beta}(k)=2(\cos{k_1}-\cos{k_2})$,
and
$g_{\gamma}(k)=-2(\sin{k_1}+\sin{k_2})$ (see Fig.~\ref{fig:g-theta}).
These mean that 
$\zeta_\alpha^\dagger$ and $\zeta_\beta^\dagger$
correspond to anisotropic spin-0 pairs  
while $\zeta_\gamma^\dagger$ corresponds to
an anisotropic spin-1 pair.

We find {from} \eqref{eq:ax} and \eqref{eq:bx} that
$\Hint$ is expressed in the wave space as
\begin{eqnarray}
\Hint  =  -\frac{1}{|\La|}
  \sum_{k,k^\prime,q\in\calK}
 W_{k,k^\prime,q}^\theta 
	    \bcupd{k+q}\bcdnd{k^\prime-q}\bcdn{k^\prime}\bcup{k}\\ 
W_{k,k^\prime,q}^\theta
      =  W(g(k+q)g_\theta(k^\prime-q)g_\theta(k^\prime)g(k)
            \nonumber\\
           +g_\theta (k+q)g(k^\prime-q)g(k^\prime)g_\theta(k)).
\end{eqnarray}
One notices that our interaction Hamiltonian expressed as above
contains scattering processes of electron pairs with non-zero total
momentum.
It should be also noted that
for scattering processes 
with zero total momentum,
only which are discussed in mean-field-type arguments,
the amplitudes   
$W_{k,-k,q}^\theta=2Wg(k+q)g_\theta(k+q)g_\theta(k)g(k)$
become either positive or negative,
depending on values of $q$ and $k$. 
Nevertheless, the ground states 
are the superpositions of products of 
the electron pairs with zero total momentum.

In the case of $\theta=\gamma$, if we consider a Hamiltonian $H_\gamma^\prime$
obtained by replacing $H_{\mathrm{int},\gamma}$ with
$H_{\mathrm{int},\gamma}^\prime
=-W\sum_{x\in\La}\sumsigma
 b_{\gamma,x,\sigma}^\dagger b_{\gamma,x,\sigma}
 \asd{x}\as{x}$,
which is interpreted as attraction between electrons with the same spin,
the following states become ground states for $\lambda\ne0$ and $W=t/4$:
$\Phi_{\gamma,\Neup,\Nedn}
=(\zeta_{\gamma,\up}^\dagger)^\frac{\Neup}{2}
                    (\zeta_{\gamma,\dn}^\dagger)^\frac{\Nedn}{2}\Phi_0$
with even positive integers $\Neup$ and $\Nedn$ less than $|\La|$,  
where the pairing operators are given by  
$\zeta_{\gamma,\sigma}^\dagger
=\sum_{x,y\in\La}(U_\gamma)_{x,y}\tasd{x}\tasd{y}$~\cite{comment3}.
For $\Neup\ne\Nedn$, $\Phi_{\gamma,\Neup,\Nedn}$ has a finite value 
of $(\Neup-\Nedn)/2$, an eigenvalue of 
the third component of the total spin.
This means that the coexistence of ferromagnetism and
spin-1 pair condensation is realized 
in the ground states of $H_{\gamma}^\prime$.
It is noted that the fully-polarized pairing states
$\Phi_{\gamma,\Neup,0}$ and $\Phi_{\gamma,0,\Nedn}$  
are stable for the on-site repulsion 
or the ferromagnetic interaction. 
These results may have some relevance to recently 
discovered materials exhibiting the superconductivity 
as well as the ferromagnetism~\cite{Saxena00,Aoki01}.
 
In the following, 
we shall prove the theorem for $\theta=\beta,\gamma$.
The case of $\theta=\alpha$ can be proved 
in a similar but slightly simple way. 

\textit{Proof of Theorem for $\theta=\beta,\gamma$.} 
We first note that, by using the $a$-operator, $\Hhop$ is rewritten
as $\Hhop=t\sum_{x}\sum_{\sigma}\asd{x}\as{x}$.
Then, using this representation of $\Hhop$ 
as well as $W=t/4$, we obtain 
\begin{eqnarray}
\lefteqn{
 \Htheta=W\sum_{x\in\La}\sumsigma 
              \asd{x}\bms{x}\bmsd{x}\as{x}}
	      \nonumber\\
 &&\hspace*{4cm} + v_\theta\sumsigma\bcsd{0}\bcs{0}.
\label{eq:Hamiltonian-2}
\end{eqnarray}
Since all the operators in the right hand side 
are positive semidefinite, 
a state which is annihilated by these operators
is a ground state, having zero energy.
We show that this is the case 
for $\PhiG$ in \eqref{eq:ground-state}.  

It follows {from} \eqref{eq:anticommutation-a}, 
\eqref{eq:expansion-b} and $(\bsd{x})^2=0$ that
\begin{eqnarray}
 \bdnd{x}\aup{x}\zthd &=& \bdnd{x}
                       \left(
			\sum_{y\in\La}
			\Uth{x,y}\tadnd{y}
			+\zthd\aup{x}
                        \right) 
                       \nonumber\\
                      &=& \zthd\bdnd{x}\aup{x}.                      
\end{eqnarray}
Noting that $(U_\beta)_{x,y}$ and $(U_\gamma)_{x,y}$ 
are symmetric and antisymmetric,
respectively, with respect to the exchange of $x$ and $y$,
we similarly obtain 
$\bupd{x}\adn{x}\zthd=\zthd\bupd{x}\adn{x}$.
These relations imply that $\PhiG$ is a zero-energy state
of the first term 
in the right hand side of \eqref{eq:Hamiltonian-2}.
Furthermore, using 
$\bcs{0}=1/\sqrt{|\La|}\sum_{x\in\La}(1-4\la)^{-1}\as{x}$
and 
\begin{equation}
\label{eq:sum-b}
 \sum_{x\in\La}\bsd{x}=0, 
\end{equation}
which follow {from}
straightforward calculations, 
we find that $\bcs{0}\zthd=\zthd\bcs{0}$. 
This together with the above result leads to $\Htheta\PhiG=0$.
Therefore, $\PhiG$ is a ground state of $\Htheta$. 
To see that    
$\PhiG$ is actually a non-zero state,
one rewrites $\zthd$ as 
\begin{equation}
 \label{eq:zthd-2}
 \zthd
  =\sum_{x\in\La\backslash\{0\}}(\taupd{x}-\taupd{0})\bdnd{x}
\end{equation}
by use of $\eqref{eq:sum-b}$. 
Since each set of 
$\{(\tasd{x}-\tasd{0})\Phi_0\}_{x\in\La\backslash\{0\}}$
and  
$\{\bsd{x}\Phi_0\}_{x\in\La\backslash\{0\}}$
is linearly independent~\cite{comment2},
$\PhiG$ is non-vanishing.

The representation \eqref{eq:zthd-2} of $\zthd$ 
motivates us to introduce the following lemma,
{from} which  
the other statements in the theorem  
follow. 

\textit{Lemma.
Suppose $\lambda\ne0$.
Any zero-energy state of $\Htheta$ with $\theta=\beta,\gamma$, 
$W=t/4$ and $\Ne$ 
less than $2|\La|$ (where $\Ne$ is not fixed) is expanded as
\begin{equation}
\label{eq:ground-state-in-lemma}
 \sum_{A\subset\La\backslash\{0\}}
   \phi_A\left(\prod_{x\in A}(\taupd{x}-\taupd{0})\right)
         \left(\prod_{x\in A}\bdnd{x}\right)\Phi_0
\end{equation}
where the coefficients $\phi_A$ satisfy
$\phi_A=\phi_{A^\prime}$ for any subsets 
$A,A^\prime$ such that $|A|=|A^\prime|$.}

This lemma implies that the ground state energy 
for odd $\Ne$ is positive.
Suppose that there are 
two linearly independent zero-energy states
for fixed even $\Ne$.
Since both of these states must satisfy 
the statement in the lemma, 
we find that the one is always
represented by the other, 
which contradicts the assumption.   
Therefore, the ground state 
for fixed even $\Ne$ is unique.
\qed

\textit{Proof of Lemma.}
The parameter $\theta$ is assumed to be $\beta$ or $\gamma$ in this proof.
Let us define $\tasp{0}=\bcs{0}$ and $\tasp{x}=\tas{x}-\tas{0}$ 
for $x\in\La\backslash\{0\}$ and 
also define $\bsp{0}=\bcs{0}$ and $\bsp{x}=\bs{x}$ 
for $x\in\La\backslash\{0\}$.
These new operators satisfy the anticommutation relations
\begin{equation}
\label{eq:anticommutation-prime}
\{\taspd{0},\tasp{x}\}
=\{\bspd{0},\bsp{x}\}=\delta_{0,x} 
\end{equation}
for $x\in\La$.
Furthermore, each set of 
$\{\taspd{x}\Phi_0\}_{x\in\La}$ and
$\{\bspd{x}\Phi_0\}_{x\in\La}$ 
is linearly independent and spans 
the single-electron Hilbert space.
Thus, the collection of states
$
\Phi_{(A,B)}^\upsilon
=\left( \prod_{x\in A}\tilde{a}_{x,\upsilon}^{\prime\dagger} \right) 
 \left( \prod_{x\in B}b_{\theta,x,-\upsilon}^{\prime\dagger}\right)\Phi_0
$
with subsets $A$ and $B$ such that $|A|+|B|=\Ne$ 
forms a complete basis for the $\Ne$-electron Hilbert space.
Here, the spin index $\upsilon$ is fixed to either $\up$ or $\dn$.  

Let $\Phi$ be an arbitrary zero-energy state of $\Htheta$ with $W=t/4$.
We first expand $\Phi$ in terms of the basis states $\Phi_{(A,B)}^\dn$ as
$
 \Phi=\sum_{A,B\subset \La}\phi_{(A,B)}\Phi_{(A,B)}^\dn
$
with coefficients $\phi_{(A,B)}$.
To be a zero-energy state,
$\Phi$ must satisfy $\bcs{0}\Phi=0$ and $\bmsd{x}\as{x}\Phi=0$
for $\sigma=\up,\dn$ and $x\in\La$.
{From} the former condition and 
\eqref{eq:anticommutation-prime},
we find that $\phi_{(A,B)}=0$ if 0 is contained in 
either $A$ or $B$, or both.
{From} the latter condition for $x_0\in\La\backslash\{0\}$ 
with $\sigma=\dn$ and 
$\{ \taspd{x},\as{y} \} = \delta_{x,y}$
for $x,y\in\La\backslash\{0\}$, 
we obtain
\begin{eqnarray}
\lefteqn{
 \sum_{A,B\subset\La\backslash\{0\}}
\chi[x_0\in A,x_0\notin B]\sgn[x_0;A,B]
}\nonumber\\
&&
\hspace*{2cm}\times
\phi_{(A,B)}\Phi_{(A\backslash\{x_0\},B\cup\{x_0\})}^\dn=0,
\end{eqnarray}
where $\sgn[\cdots]$ is a sign factor coming {from} exchanges of
the fermion operators, and $\chi[\textrm{``event''}]$ takes 1
if ``event'' is true and 0 otherwise.
Since all the terms in the left hand side 
are linearly independent, we find $\phi_{(A,B)}=0$ 
if $x_0\in A$ in addition to $x_0\notin B$.
This holds for any $x\in\La\backslash\{0\}$, so that
only the terms with $A,B$ such that 
$A\subset B\subset\La\backslash\{0\}$ can contribute to the expansion. 
Taking account of the above results, we rewrite $\Phi$ as
$
\Phi=\sum_{A,B\subset\La;|A|\ge|B|}\phi_{(A,B)}^\prime\Phi_{(A,B)}^\up
$
with new coefficients $\phi_{(A,B)}^\prime$.
Operating $\bcs{0}$ and $\bdnd{x_0}\aup{x_0}$ on $\Phi$
in this form and
repeating a similar argument to the above, we find that
$\Phi$ is expanded in
terms of $\Phi_{(A,B)}^\up$ 
with $A,B$ such that $A=B\subset \La\backslash\{0\}$.

Any zero-energy state is thus written as
$
 \sum_{A\subset\La\backslash\{0\}}\phi_A\Phi_{(A,A)}^\up
$
where $\phi_A=\phi_{(A,A)}^\prime$.
Then, we again consider the zero-energy state condition
$\bupd{x_0}\adn{x_0}\Phi=0$
for $x_0\in\La\backslash\{0\}$
and derive conditions on $\phi_A$.
Here, it is noted that $\bs{x_0}$ is expanded as
$\bs{x_0}=\sum_{y\in\La\backslash\{0\}}\Uth{y,x_0}\tasp{y}$.
{From} this and the anticommutation relation 
$\{\bspd{x},\as{x_0}\}=\{\bsd{x},\as{x_0}\}=\Uth{x,x_0}$ 
for $x\in\La\backslash\{0\}$, we deduce
\begin{eqnarray}
\lefteqn{
 \sum_{A\subset\La\backslash\{0\}}
  \sum_{y,y^\prime\in\La\backslash\{0\}}
  \chi[y\notin A, y^\prime\in A] \sgn[y,y^\prime;A]
}\nonumber\\
&&\hspace*{2cm}
\times
F_{y,y^\prime}^{x_0}\phi_A\Phi_{(A\cup\{y\},A\backslash\{y^\prime\})}^\up = 0
\end{eqnarray}
where $F_{y,y^\prime}^{x_0}=\Uth{x_0,y}\Uth{y^\prime,x_0}$, and
$\sgn[\cdots]$ is a fermion sign factor.
Let us choose a subset $A$ which does not contain 
a nearest-neighbor site $y$ of $x_0$ 
but does contain next-nearest-neighbor
site $y^\prime$ in the same direction.
(The sites $x_0$, $y$ and $y^\prime$ are in the same axis.)
For this set of sites, $F_{y,y^\prime}^{x_0}$ is non-zero.
Then, by checking the coefficient of 
$\Phi_{(A\cup\{y\},A\backslash\{y^\prime\})}^\up$,
we have 
$
(\sgn[y,y^\prime;A]\phi_A
 +\sgn[y^\prime,y;A_{y^\prime\to y}]\phi_{A_{y^\prime\to y}})=0
$
where $A_{x\to y}$ 
is defined for $x\in A$ and $y\notin A$ 
by $A_{x\to y}=(A\backslash\{x\})\cup \{y\}$.  
Since 
$\sgn[y,y^\prime;A]
 =-\sgn[y^\prime,y;A_{y^\prime\to y}]$,
we obtain ${\phi_{A}=\phi_{A_{y^\prime\to y}}}$.

Repeating the same argument for all $x\in\La\backslash\{0\}$,
we reach the conclusion 
that $\phi_A=\phi_{A^\prime}$ whenever $|A|=|A^\prime|$,
which completes the proof of the lemma. 
\qed  

Now let us consider the case of $\lambda=0$ at half-filling with 
the inclusion of the on-site repulsion.
Here, we furthermore assume 
that $v_\theta=0$ for all $\theta$ and 
that $L_1$ and $L_2$ are even integers.
In this case the Hamiltonian becomes 
$H_{\theta,U}^{\lambda=0}
=W\sum_{x}\sum_{\sigma}\csd{x}\bms{x}\bmsd{x}\cs{x}
+U\sum_{x}c_{x,\up}^\dagger c_{x,\dn}^\dagger c_{x,\dn}c_{x,\up}$ with $U>0$,
which is still positive semidefinite. 
At half-filling, zero-energy states 
of on-site repulsion term are given by 
$\left(\prod_{x\in\La}c_{x,\sigma_x}^\dagger\right)\Phi_0$
with $\sigma_x=\up,\dn$.
Then, considering the zero-energy condition
for the first term in the Hamiltonian, 
we conclude that the ground states of $H_{\theta,U}^{\lambda=0}$
are two-fold degenerate and given by the N\'eel states, 
exhibiting antiferromagnetism.

One can readily find 
that the Hamiltonian $\Htheta$ does not possess 
spin rotational symmetry.
In the case of $\theta=\alpha,\beta$, however,
we can construct an isotropic model with the ground
state~\eqref{eq:ground-state} as follows.
Let us define
$
 H_{\mathrm{int,\theta}}^\prime
  =\frac{W}{2}\sum_{x}
  (\aupd{x}\bup{x}-\adnd{x}\bdn{x})
  (\bupd{x}\aup{x}-\bdnd{x}\adn{x})
$ for $\theta=\alpha,\beta$.
A straightforward  
calculation yields that 
$(\bupd{x}\aup{x}-\bdnd{x}\adn{x})\zthd
=\zthd(\bupd{x}\aup{x}-\bdnd{x}\adn{x})$~\cite{comment5}, and, 
since 
$H_{\mathrm{int,\theta}}^\prime$ is positive semidefinite,
\eqref{eq:ground-state} remains 
the ground state of $\Htheta+H_{\mathrm{int,\theta}}^\prime$ with
$W=t/4$
for $\theta=\alpha,\beta$.
Furthermore, 
$\Htheta+H_{\mathrm{int,\theta}}^\prime$ is isotropic
since it
commutes with $S_\mathrm{tot}^{(3)}
=\sum_{x}(c_{x,\up}^\dagger c_{x,\up} - c_{x,\dn}^\dagger c_{x,\dn})/2$
and  
$S_\mathrm{tot}^+=\sum_{x}c_{x,\up}^\dagger c_{x,\dn}$~\cite{comment6}.  
A construction of an isotropic model 
for the spin-1 pairing case and 
detailed investigation of 
perturbed models of ours in both spin-0 and 1 cases  
are left as an interesting future study.        


\begin{thebibliography}{99}
\bibitem{comment4}
	The uniqueness of the ground state 
	excludes an appearance of   
	physically less interesting phenomena 
	such as the paramagnetism and the translational 
	symmetry breaking in a finite system.
	It guarantees that one can discuss the physical properties
	of the model by analyzing the exact ground states.
\bibitem{comment1}
	Here $|\cdot|$ denotes the Euclidean norm. 
	The same symbol $|X|$ is used to denote
	the number of elements in a set $X$.
\bibitem{comment3}
	{From} the wave-space representation of 
	$\zeta_{\gamma,\sigma}^\dagger$,
	one finds that the states $\Phi_{\gamma,\Neup,\Nedn}$ are 
	non-vanishing.
	Unfortunately, the uniqueness of the ground state
	for fixed values of $N_{\mathrm{e},\up}$ and 
	$N_{\mathrm{e},\dn}$ is not proved at present.
\bibitem{comment2}
	The linear independence of 
	$\{\bsd{x}\Phi_0\}_{x\in\La\backslash\{0\}}$ 
	with $\theta=\beta,\gamma$
	is proved as follows. Let $G_\theta$ be 
	the Gram matrices of 
	$b$-operators whose matrix elements are given by
	$(G_{\theta})_{x,y}=\{\bsd{x},\bs{y}\}$.
	The eigenvalues of $G_\theta$ are 
	$g_\theta^2(k)$ with $k\in\calK$.
	Since $L_1$ and $L_2$ are odd integers and differ by 2, 
	$g_\theta(k)$ with $\theta=\beta,\gamma$ become zero
	if and only if $k=(0,0)$.
	Thus, the dimension of the kernel of 
	$G_\theta$ with $\theta=\beta,\gamma$ is one.
	This together with \eqref{eq:sum-b} proves the linear independence.
	In the case of $\theta=\alpha$, all $g_\alpha(k)$ are non-zero,
	so that $\{b_{\alpha,x,\sigma}^\dagger\Phi_0\}_{x\in\La}$ 
	is linearly independent.
\bibitem{comment6} 
        We note that $\Hint+H_{\mathrm{int},\theta}^\prime$ is rewritten 
	as  
        $\frac{2W}{t}\Hhop+W\sum_x\{
        {\mbox{\boldmath $S$}_{x}^a \cdot \mbox{\boldmath $S$}_{x}^b}
        -\frac{3}{4}n_{x}^a n_{x}^b
        \}$
        where 
        $n_{x}^a = \sum_{\sigma} a_{x,\sigma}^\dagger a_{x,\sigma}$,
        $(S_{x}^{a})^{(l)}=$ 
        $\sum_{\sigma,\tau}
         a_{x,\sigma}^\dagger(p_{\sigma,\tau}^{(l)}/2) a_{x,\tau}$ 
        with the Pauli matrices $p^{(l)}$, and $n_{x}^b$ etc. are 
        defined similarly for the $b$-operator.
\end{thebibliography}
\end{document}